\documentclass[%
twocolumn,
 superscriptaddress,
 amsmath,amssymb,
 aps,
 prmaterials,
]{revtex4-2}

\usepackage{graphicx,subfigure}
\usepackage{dcolumn}
\usepackage{bm}
\usepackage{float}
\usepackage{xcolor}
\usepackage{siunitx} 
\usepackage[version=4]{mhchem} 
\usepackage{bbm} 
\usepackage{lipsum}
\usepackage{lineno}

\newcommand{\matr}[1]{\bm{#1}}

\DeclareSIUnit{\formulaunit}{{f.u.}}

\newcommand\numberthis{\addtocounter{equation}{1}\tag{\theequation}}

\newlength{\figurewidth}
\newlength{\doublefigurewidth}
\begin{document}


\title{Charting the thermodynamic stability of hybrid perovskite alloys with machine learning}


\author{Jarno Laakso}
\affiliation{%
 Department of Applied Physics, Aalto University, Espoo, Finland
}%

\author{Armi Tiihonen}
\affiliation{%
 Department of Mechanical Engineering, Chalmers University of Technology, Gothenburg, Sweden
} 

\author{Patrick Rinke}
\email{patrick.rinke@tum.de}
\affiliation{%
 Department of Applied Physics, Aalto University, Espoo, Finland
}%
\affiliation{Physics Department, Technical University of Munich, Garching, Germany}
\affiliation{Atomistic Modelling Center, Munich Data Science Institute, Technical University of Munich, Garching, Germany}
\affiliation{Munich Center for Machine Learning (MCML)}

\date{\today}

\begin{abstract}
Alloy-based perovskite semiconductor materials offer tunable properties and improved stability, but their complexity has impeded accurate modeling, hindering development of solar cell devices. We present a machine-learning (ML) accelerated atomistic modeling approach for the phase stability of \ce{(Cs/FA)Pb(Br/I)3} and \ce{(Cs/FA)Sn(Br/I)3} perovskites (where FA is formamidinium), that explicitly accounts for finite-temperature entropy effects. To make such quaternary alloy calculations tractable, we adopt a two-level strategy, combining 1) graph neural network interatomic potentials trained on density functional theory data for efficient structure relaxations with 2) direct energy predictions from unrelaxed structures to further accelerate the process. Our strategy enables computations of free energy landscapes across compositions and phases, capturing alloy disorder and molecular orientations of FA. Our results reveal narrower regions of stable compositions for the Sn-based system compared to its Pb-based counterpart, limiting compositional engineering strategies for realizing devices that are robust in long term. High I contents result in maximum stability and no stabilization is observed near the center of the composition space. Our results guide the design of stable perovskite-based devices in experiments and demonstrate a workflow that is generalizable to other complex alloy modeling tasks.
\end{abstract}

\maketitle

Perovskite photovoltaics research is advancing rapidly and holds the promise of even lower energy payback times compared to the current commercially available solar cell technologies \cite{gong2015perovskite, xueyu2020life, tian2021life}. The most efficient perovskite solar cells (PSCs) have already surpassed power conversion efficiency of 26\% \cite{liu2023bimolecularly, liang2023homogenizing}, which rivals the leading silicon-based alternatives. Moreover, tandem cells that pair perovskites with silicon provide even higher efficiencies \cite{liu2024perovskite}. Despite notable advancements, the commercial deployment of PSCs continues to face significant obstacles, primarily due to unresolved issues surrounding long-term operational stability \cite{khenkin2020consensus, chowdhury2023stability} and concerns over toxic lead present in the most efficient perovskite compositions \cite{ke2019prospects, zhang2021lead}.

The properties of perovskite materials can be systematically tailored through compositional engineering \cite{saliba2019polyelemental}. Among these, alloys derived from the hybrid perovskite \ce{FAPbI3} (FA = formamidinium, \ce{CH(NH2)2+}) have emerged as promising candidates for enhancing both the stability and environmental compatibility of PSCs \cite{liang2024toward}. Compositions \ce{(Cs/FA)$B$(Br/I)3} are frequently considered, where partial substitution of FA with Cs has improved structural and thermal stability \cite{sun2021data}, halide mixing at the $X$-site enables precise band gap tuning \cite{berger2023how}, and replacement of Pb with Sn at the $B$-site offers a pathway to reduced toxicity. Despite these advances, the fundamental mechanisms governing the stability of such quaternary alloys remain incompletely understood.

The \ce{(Cs/FA)Pb(Br/I)3} alloys have so far received the most attention \cite{mcmeekin2016mixed, marchezi2020degradation, wang2023sustainable, hidalgo2024br}. Wang et al., for example, conducted the most comprehensive experimental exploration of the two-dimensional compositional space, synthesizing and characterizing 81 distinct compositions \cite{wang2023sustainable}. In contrast, the compositions of the lead-free counterpart \ce{(Cs/FA)Sn(Br/I)3} remain comparatively underexplored. To date, experimental investigations have been limited to binary systems such as \ce{(Cs/FA)SnI3}, \ce{FASn(Br/I)3}, and \ce{CsSn(Br/I)3} \cite{gao2018robust, pansa-ngat2021phase, berger2023how, park2023compositional}, leaving much of the quaternary compositional landscape uncharted.

Computational materials science offers a powerful route to accelerate the discovery of improved perovskite materials by elucidating the fundamental mechanisms underlying their intrinsic instabilities and enabling predictions for compositions yet to be explored experimentally. However, traditional computational methods face limitations due to the inherent complexity of these multicomponent systems. While density functional theory (DFT), the cornerstone of atomistic modeling, provides accurate insight at the electronic and structural levels, its application becomes computationally prohibitive when navigating the vast configurational space of multicomponent perovskite alloys.

\begin{figure*}
    \centering
    \includegraphics[width=\doublefigurewidth]{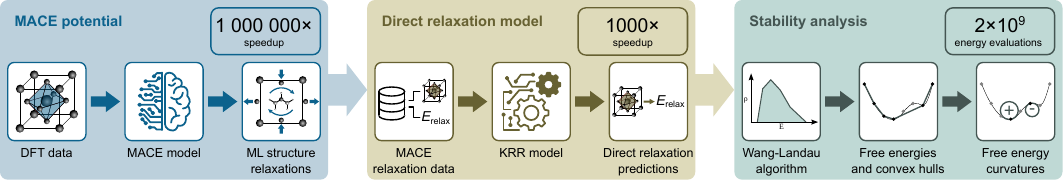}
    \caption{Computational workflow for model training and stability analysis. First, a MACE interatomic potential is trained on DFT data. Then, Kernel ridge regression (KRR) models are trained based on the MACE relaxation data to predict relaxed energies directly from the unrelaxed atomic structures. Finally, the direct relaxation models are employed in free energy calculations of perovskite alloys, providing information on stability.}
    \label{fig:fig_1}
\end{figure*}

To address the computational cost of DFT in alloy studies, cluster expansions have been widely used to rapidly predict energies by fitting models to a limited set of DFT calculations \cite{sanchez1984generalized}. The cluster expansion has proven effective for analyzing the stability of binary halide perovskite alloys with $X$-site \cite{yamamoto2017structural, bechtel2018first} or $B$-site \cite{maleka2024phase} substitutions. However, as an on-lattice approach, it cannot capture the distinct orientations of molecular cations in hybrid perovskites. An alternative is the special quasirandom structures (SQS) method \cite{zunger1990special}, which approximates alloy properties using a small number of representative configurations. SQS has been applied, for example, to compute Gibbs free energy curves for binary hybrid perovskites such as \ce{(Cs/FA)PbI3} and \ce{(FA/MA)PbI3} \cite{schelhas2019insights}. Yet, it lacks an inherent treatment of finite-temperature effects, and entropy contributions are typically introduced via analytical approximations. Moreover, SQS's reliance on a few configurations introduces significant biases when modeling complex alloy behavior.

Machine learning (ML) offers a powerful alternative to traditional alloy approaches by enabling accurate predictions for structurally complex materials at a fraction of DFT’s computational cost. Atomistic ML models are now widely used for both inorganic \cite{baldwin2023dynamic, thomas2019machine} and hybrid \cite{chen2019fast, jinnouchi2019phase, bokdam2021exploring, karimitari2024accurate} single-component perovskites, with recent extensions to alloy systems. In prior work, we introduced a method using ML-based structure relaxations to sample alloy convex hulls, applying it to identify stable compositions in \ce{CsPb(Br/I)3} \cite{laakso2022compositional}. Other studies have followed similar paths: one used graph neural networks to model \ce{CsPbI3} with Cd, Zn, and Br substitutions \cite{krautsou2025impact}, while others have employed neural network potentials to simulate phase segregation in \ce{(MA/FA)Pb(Br/I)3} \cite{chen2021microstructure} and \ce{(Cs/FA)Pb(Br/I)3} \cite{luo2025machine}. However, these approaches generally neglect finite-temperature entropy effects, limiting their ability to fully capture thermodynamic stability.

In this study, we explore the thermodynamic phase stability of the hybrid perovskite alloys \ce{(Cs/FA)Pb(Br/I)3} and \ce{(Cs/FA)Sn(Br/I)3} by constructing their free energy landscapes using ML (workflow in Fig. \ref{fig:fig_1}). Building on our previous work for inorganic perovskites, we extend our ML framework to hybrid systems, where orientable organic cations introduce added complexity. We train MACE interatomic potentials \cite{batatia2022mace, batatia2022design} on DFT data to enable efficient structure relaxations, supported by an optimized training workflow developed earlier \cite{homm2025efficient}. Finite-temperature effects are incorporated via the Wang-Landau algorithm \cite{wang2001efficient}. To accelerate the extensive sampling required, we introduce secondary ML models trained on MACE-relaxed structures to predict relaxed energies directly from unrelaxed configurations. The workflow is rigorously validated at each stage and benchmarked against available experimental data \cite{wang2023sustainable, gao2018robust, pansa-ngat2021phase}.

Our work is guided by the hypothesis that the reduced thermodynamic stability of tin-based hybrid perovskites arises from fundamental differences in their free energy landscapes compared to their lead-based counterparts. Accordingly, our primary objective is to compare the thermodynamic behavior of \ce{(Cs/FA)Pb(Br/I)3} and \ce{(Cs/FA)Sn(Br/I)3} to uncover the origins of this disparity. In doing so, we also aim to address the broader question of whether entropy-driven stabilization can enhance the thermodynamic stability of hybrid perovskite alloys.

\section{Results}
Our solution to the long-standing sampling challenge that has prevented computational investigations of quaternary hybrid perovskite alloys in the past follows a multi-stage workflow, summarized in Fig. \ref{fig:fig_1}. We first accelerated perovskite structure relaxations by training MACE graph neural network interatomic potentials \cite{batatia2022mace, batatia2022design} on DFT data. Although this approach speeds up computations by approximately six orders of magnitude compared to DFT, MACE-based relaxations alone are not fast enough for free energy calculations. They did, however, allow us to generate extensive relaxation data that we then used to fit a secondary set of ML models. These models employ Kernel Ridge regression (KRR) to map unrelaxed atomic structures directly to relaxed energy values, bypassing the need for explicit structure relaxation. This second step accelerated the computations by an additional three orders of magnitude, which finally allowed us to perform the energy sampling necessary for the free energy calculations with the Wang-Landau algorithm \cite{wang2001efficient}.

In this section, we first present results from model training, assessing the accuracy of structural relaxations with the fitted MACE potentials, and comparing the predictions of the direct relaxation models to MACE-based relaxations. Then, we present the computed free energy landscapes for various perovskite phases of both \ce{(Cs/FA)Pb(Br/I)3} and \ce{(Cs/FA)Sn(Br/I)3}, identifying energetically favorable regions of the composition spaces. We also show the convex hulls and curvatures of the free energy surfaces to analyze thermodynamic stability against phase separation.

\begin{figure}
    \centering
    \includegraphics[width=\figurewidth]{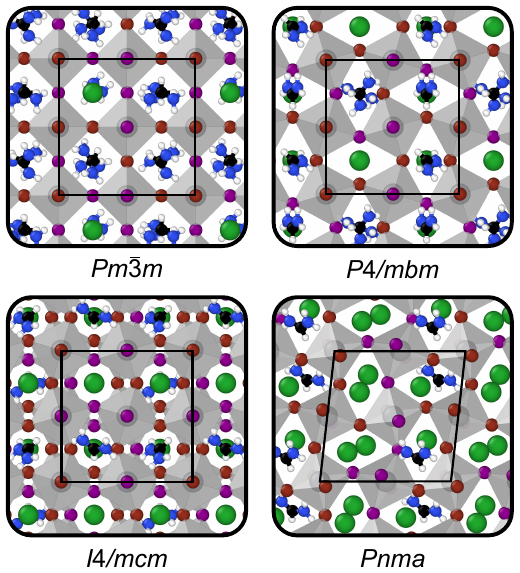}
    \caption{Example atomic structures of the four perovskite phases of \ce{(Cs/FA)$B$(Br/I)3}
    investigated in this work. The solid lines indicate the 2$\times$2$\times$2 supercells, containing carbon (black), nitrogen (blue), hydrogen (white), cesium (green) lead/tin (gray),  bromide (brown), and iodine (purple) atoms.}
    \label{fig:fig_2}
\end{figure}

\subsection{Model training}
The MACE potentials were trained on DFT calculations to accelerate structure relaxations, following the efficient data generation workflow we developed earlier \cite{homm2025efficient} (see Methods for details). The final models were fitted using training sets comprising \num{2000} DFT calculations for both the Pb- and Sn-based alloys. The data included $2\times 2\times 2$ perovskite supercell structures from four different phases: $Pm\overline{3}m,$ $P4/mbm,$ $I4/mcm,$ and $Pnma$ (illustrated in Fig. \ref{fig:fig_2}). 

\begin{figure}
    \centering
    \includegraphics[width=\figurewidth]{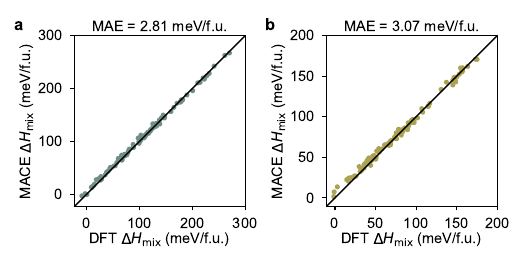}
    \caption{Test set relaxation energy predictions of the trained MACE models vs. the corresponding DFT values for (a) \ce{(Cs/FA)Pb(Br/I)3} and (b) \ce{(Cs/FA)Sn(Br/I)3} with the mean absolute errors over the whole test datasets.
    }
    \label{fig:fig_3}
\end{figure}

Ideally, the relaxation accuracy of these models would be evaluated by relaxing a representative set of structures using both DFT and MACE, followed by a direct comparison of the results. However, this approach was not feasible due to the substantial computational cost associated with DFT relaxations -- performing them on a statistically meaningful test set would require an order of magnitude more DFT calculations than those used for training the MACE model. Instead, we evaluated the relaxation accuracy by relaxing 100 random alloy configurations with both MACE models. We then performed single-point DFT calculations on the MACE-relaxed structures and compared the DFT energies to the MACE predictions. The corresponding mean absolute errors (MAEs) are listed in the first column of Table \ref{tab:tab_1} for each phase. The overall MAEs per perovskite formula unit (f.u.) are \qty{2.81}{meV/\formulaunit} and \qty{3.07}{meV/\formulaunit} for \ce{(Cs/FA)Pb(Br/I)3} and \ce{(Cs/FA)Sn(Br/I)3}, respectively. Visualization of the same comparison (Fig. \ref{fig:fig_3}) shows that the model accuracy for the relaxed structures remains consistent across the whole energy range.

Next we trained the direct relaxations models on MACE relaxation data. To assess their accuracy, an initial test was conducted following initial data generation. The initial datasets -- consisting of relaxation data obtained through clustering and Monte Carlo (MC) sampling -- were divided into 80\% training and 20\% test data. The resulting prediction MAEs on the test set, reported in the second column of Table \ref{tab:tab_1}, were \qty{10.0}{meV/\formulaunit} on average, which is considerably higher than those observed for the MACE models. 

To ensure reliable free energy calculations, it was essential to improve the accuracy of the direct relaxation models, especially in the low-energy regime. To achieve this, we applied an active learning scheme that iteratively reduces prediction errors on low-energy alloy configurations through repeated energy minimization MC simulations. To asses the final models, we utilized them in performing energy minimization simulations at all alloy compositions permitted by the 2$\times$2$\times$2 supercell and compared their energy predictions on the obtained minimum alloy configurations to the corresponding MACE relaxation energies. To guarantee unbiased test results, we made sure that no minimum-energy configurations included in the training sets of the models were used in the evaluation. The results of the test (third column of Table \ref{tab:tab_1}) reveal that the MAEs of the direct relaxation models on minimum-energy configurations now average \qty{3.2}{meV/\formulaunit} -- a considerable improvement compared to the initial errors.

\begin{table}[]
    \centering
    \begin{tabular}{lc@{\hskip 2mm}c@{\hskip 2mm}c}
        \hline\vspace{-2mm}\\
            & MAE $E_{\text{relax}}^{\text{MACE}}$ & MAE $E^{\text{KRR}}$ & MAE $E_{\text{min}}^{\text{KRR}}$\\
            & (meV/f.u.) & (meV/f.u.) & (meV/f.u.) \\
            \hline
            \ce{(Cs/FA)Pb(Br/I)3} & & & \\
            \hspace{3mm}$Pm\overline{3}m$             & 3.04 & 6.15  & 1.78 \\
            \hspace{3mm}$P4/mbm$                      & 2.45 & 12.16 & 2.93 \\
            \hspace{3mm}$I4/mcm$                      & 1.96 & 10.41 & 4.29 \\
            \hspace{3mm}$Pnma$                        & 3.78 & 12.20 & 3.84 \\
            \hspace{3mm}\textbf{Overall}\vspace{2mm}  & 2.81 & 10.23 & 3.21 \\
            \ce{(Cs/FA)Sn(Br/I)3} & & & \\
            \hspace{3mm}$Pm\overline{3}m$             & 2.68 & 7.07  & 2.36 \\
            \hspace{3mm}$P4/mbm$                      & 2.57 & 10.95 & 2.79 \\
            \hspace{3mm}$I4/mcm$                      & 2.82 & 10.78 & 3.82 \\
            \hspace{3mm}$Pnma$                        & 4.22 & 10.18 & 3.59 \\
            \hspace{3mm}\textbf{Overall}              & 3.07 & 9.75  & 3.14 \\
            \hline
    \end{tabular}
    \caption{Mean absolute errors (MAEs) of energy predictions at different stages of the ML workflow for the lead- and tin-based composition spaces. Shown are MAEs for: (1) MACE relaxations ($E_{\text{relax}}^{\text{MACE}}$), compared to single-point DFT calculations on MACE-relaxed structures; (2) direct relaxation models before active learning ($E^{\text{KRR}}$), evaluated against MACE-relaxed energies across the energy landscape; and (3) direct relaxation models after active learning ($E_{\text{min}}^{\text{KRR}}$), computed on the minimum-energy alloy configurations obtained at every composition.}
    \label{tab:tab_1}
\end{table}

\subsection{Free energy landscapes}
Free energies were computed at all compositions permitted by the 2$\times$2$\times$2 supercell we adopted using the Wang–Landau algorithm. Over 98\% of the simulations converged within \num{3000000} iterations. Among the converged runs, 1.2\% yielded clear outlier values, which were manually removed from further analysis (detailed information in Supplementary Section S2\,F). The Wang–Landau simulations were repeated with a different seed, and the mean absolute difference between the free energy values obtained from the two runs was \qty{0.37}{meV/\formulaunit} for the raw data, which reduced to \qty{0.28}{meV/\formulaunit} after outlier removal.

We assessed thermodynamic stability using the computed free energy landscapes through three complementary approaches. First, the Helmholtz free energy of mixing provides a direct measure of how energetically favorable a given alloy composition is compared to the pure, unalloyed perovskites. Lower values indicate greater stability, with negative values suggesting no tendency toward phase separation. Second, we constructed convex hulls of the free energy landscapes; compositions lying on the convex hull are predicted to be stable and resistant to decomposition. Finally, we examined the curvature of the free energy surfaces. A positive curvature indicates local convexity, which corresponds to stability against small compositional fluctuations. While all compositions on the convex hull have positive curvature, such values can also appear off the hull, suggesting potential metastability in those regions. It should be noted that the analysis presented here does not address the relative preferences over the structural phases, and thus it does not consider phase separation into other structural phases but only within a given phase.

\begin{figure}
    \centering
    \includegraphics[width=\figurewidth]{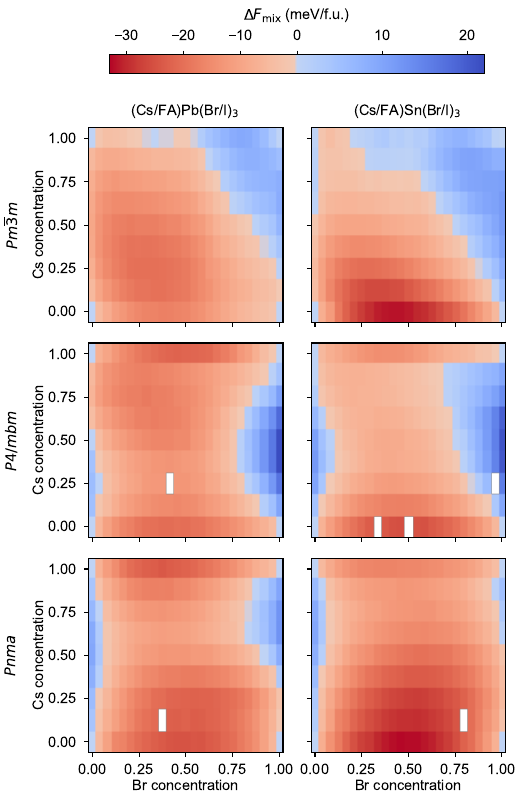}
    \caption{Helmholtz free energy of mixing ($\Delta F_{\text{mix}}$) landscapes obtained with the Wang-Landau algorithm at \qty{300}{K}. The left column contains the results for \ce{(Cs/FA)Pb(Br/I)3} and the right column for \ce{(Cs/FA)Sn(Br/I)3}. The rows correspond to three phases: cubic $Pm\overline{3}m$, tetragonal $P4/mbm$, and orthorhombic $Pnma$. White rectangles imply missing data due to convergence problems of the Wang-Landau algorithm.}
    \label{fig:fig_4}
\end{figure}

\begin{figure}
    \centering
    \includegraphics[width=\figurewidth]{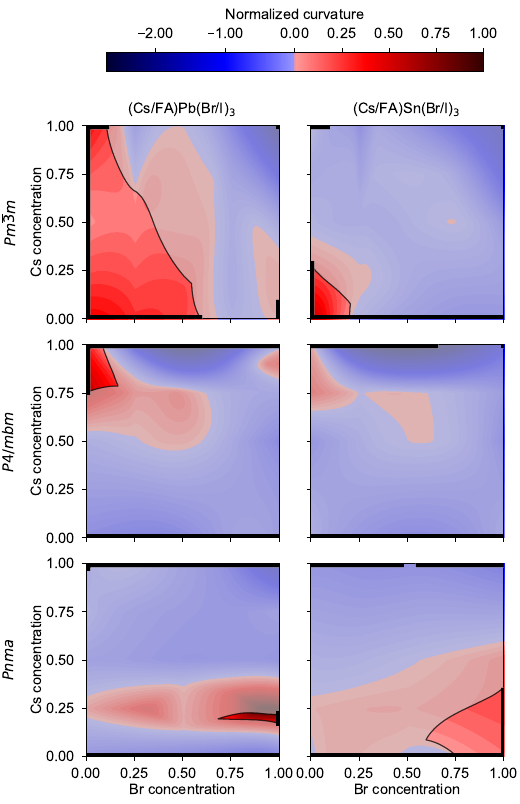}
    \caption{Curvature of the free energy surfaces at \qty{300}{K}. The curvature values have been normalized so that the highest curvature observed across the phases equals one. The compositions that are on the convex hull are highlighted by brighter red tones. Bold black lines on the edges indicate binary alloy compositions that are on the convex hull. The left column contains the results for \ce{(Cs/FA)Pb(Br/I)3} and the right column for \ce{(Cs/FA)Sn(Br/I)3}. The three rows correspond to the phases: cubic $Pm\overline{3}m$, tetragonal $P4/mbm$, and orthorhombic $Pnma$.}
    \label{fig:fig_5}
\end{figure}

Temperature affects the free energy landscapes through entropic contributions. Figures \ref{fig:fig_4} and \ref{fig:fig_5} show results obtained at \qty{300}{K}, which is a reasonable temperature for storing perovskite samples and perovskite solar cells. However, higher temperatures are frequently encountered, e.g., during synthesis. Thus, we analyzed the free energy and its curvatures at the elevated temperature of \qty{150}{\degree C} and show them, for brevity, in Supplementary Figs. S13 and S14.

The free energy landscapes, averaged over simulations using two independent random seeds, are presented in Fig.~\ref{fig:fig_4}. Results for the $I4/mcm$ phase are excluded due to inconsistencies observed in structure relaxations caused by its structural proximity to the $P4/mbm$ phase, which led to unreliable free energy values. Instead, we present the $P4/mbm$ results here as representative of the tetragonal phases, while the results for $I4/mcm$ are available in Supplementary Section S2\,G. For the three remaining phases, both \ce{(Cs/FA)Pb(Br/I)3} and \ce{(Cs/FA)Sn(Br/I)3} exhibit similar overall trends with some notable differences.

In the $Pm\overline{3}m$ phase,  \ce{(Cs/FA)Pb(Br/I)3} features a low-energy region in the middle of the 2D composition space, with a well depth of approximately \qty{20}{meV/\formulaunit} and a minimum located around 25\% Cs and 30\% Br. In contrast, for \ce{(Cs/FA)Sn(Br/I)3} the lowest energies are reached at the Cs-free edge of the composition space without the well in the middle of the composition range observed for \ce{(Cs/FA)Pb(Br/I)3}.

For the $P4/mbm$ phase, the key distinction between the Pb- and Sn-based systems lies in the behavior near the \ce{Cs$B$(Br/I)3} binary edge: in the Pb-based system, the low-energy region extends into quaternary alloys with high Cs and low Br concentrations, indicating enhanced stability. This extension, however, is absent in the Sn-based counterpart. The free energy landscapes for $Pnma$ resemble those of $P4/mbm,$ with the exception that neither material space exhibits local free energy minima in the central region of the 2D composition space.

We then analyzed the thermodynamic stability further by constructing the convex hulls and curvatures of the free energy surfaces. The convex hulls for the lead- and tin-based composition spaces are presented in Fig. \ref{fig:fig_5} by bright red areas (for quaternary alloys) and bold black lines (for binary alloys). In the $Pm\overline{3}m$ phase, both material spaces exhibit a stable region on the convex hull near the lower-left corner of the composition map -- corresponding to low Br and low Cs concentrations. For \ce{(Cs/FA)Pb(Br/I)3} this stable region is broader, spanning the whole range of $A$-site substitutions and extending up to 60\% Br concentration depending on the Cs/FA ratio. Within the $P4/mbm$ phase, only the \ce{(Cs/FA)Pb(Br/I)3} system features quaternary alloy compositions on the convex hull, specifically near the top-left corner of the composition space, corresponding to high Cs and low Br content. For $Pnma,$ \ce{(Cs/FA)Pb(Br/I)3} exhibits a small stable region at high Br content at approximately 20\% Cs concentration. Stable compositions are also present in the corresponding part of the \ce{(Cs/FA)Sn(Br/I)3} landscape, but the stable region is slightly broader, extending all the way to the corner of the composition space.

Figure \ref{fig:fig_5} also displays the curvature landscapes of the free energy surfaces on and outside the convex hull. For the $Pm\overline{3}m$ phase of \ce{(Cs/FA)Pb(Br/I)3}, the highest curvature values -- indicating the highest stability -- are located near the \ce{CsPbI3} and \ce{FAPbI3} corners of the composition space, as well as along the Cs-free edge of the composition space up to 50\% Br concentration. Positive curvature values are also observed at compositions outside the convex hull, most notably near the center and upper regions of the composition space (around 50\% Br content), indicating possible metastability. For \ce{(Cs/FA)Sn(Br/I)3}, the highest curvature is found near the \ce{FASnI3} corner -- also stable according to the convex hull analysis.

In the $P4/mbm$ phase, \ce{(Cs/FA)Pb(Br/I)3} exhibits very high curvature values in the proximity of the \ce{CsPbI3} corner. Moreover, both \ce{(Cs/FA)Pb(Br/I)3} and \ce{(Cs/FA)Sn(Br/I)3} display regions of positive curvature in the upper (Cs-rich) halves of the landscapes. In the $Pnma$ phase, this behavior is mirrored, with metastable compositions in the Pb-based alloy space forming a band at lower Cs content of approximately 25\%. For \ce{(Cs/FA)Sn(Br/I)3}, this band is broader and covers most low-Cs compositions.

\section{Discussion}
To evaluate the effectiveness of our ML approach, we first examined the performance of the MACE potentials. The models demonstrated rapid convergence, requiring remarkably few data points -- final training sets contained only \num{2000} atomic structures. The relaxation energy errors for the final MACE potentials were \qty{2.8}{meV/\formulaunit} and \qty{3.1}{meV/\formulaunit} for \ce{(Cs/FA)Pb(Br/I)3} and \ce{(Cs/FA)Sn(Br/I)3}, respectively (Fig.~\ref{fig:fig_3}). The errors translate to \qty{0.35}{meV/atom} on average, which compares favorably to previous ML potentials for hybrid perovskites \cite{bokdam2021exploring, karimitari2024accurate} as well as our earlier work on inorganic systems \cite{laakso2022compositional, homm2025efficient}. Other metrics, such as single-point energy and force predictions (Supplementary Sections S2\,A and S2\,B), were equally good, further confirming the strong suitability of MACE in modeling complex perovskite systems.

By contrast, the direct relaxation models trained on the initial datasets (obtained via clustering and MC sampling) yielded higher prediction errors: \qty{10.2}{meV/\formulaunit} for \ce{(Cs/FA)Pb(Br/I)3} and \qty{9.8}{meV/\formulaunit} for \ce{(Cs/FA)Sn(Br/I)3} (Table \ref{tab:tab_1}). Increasing the training set size did not lead to substantial improvements in accuracy (see Supplementary Section S2\,C). In our view, tailoring of the direct relaxation model architecture remains a key challenge for future methodological improvement.

Further training of the direct relaxation models with active learning improved predictions greatly in the low-energy regime (Table \ref{tab:tab_1}) -- the domain most critical for free energy calculations. Our analysis (Supplementary Section S2\,E) identifies two mechanisms behind the improvement: 1) enhanced accuracy on actual low-energy configurations, and 2) suppression of unrealistically low energy predictions for configurations that, in reality, lie much higher on the energy landscape. Without active learning, the high errors in the low-energy regime would have significantly degraded the quality of the free energy results, highlighting its essential role in the workflow.

Together, the two ML stages accelerated energy predictions by approximately nine orders of magnitude, enabling the evaluation of over two billion relaxation energies necessary to construct the free energy landscapes for both \ce{(Cs/FA)Pb(Br/I)3} and \ce{(Cs/FA)Sn(Br/I)3}. The findings of this study are intended to guide future experimental efforts in perovskite screening and compositional design. To support this goal, we evaluated how well our results align with existing experimental observations. We compared our findings for \ce{(Cs/FA)Pb(Br/I)3} to the experimental phase mapping and photostability study by Wang et al. \cite{wang2023sustainable}, which succeeded in sampling this composition space in full with a high-throughput setup. Our predictions for the cubic phase correspond well to the experiments, with the majority of the stable compounds concentrated in the low-Br, low-Cs region of the composition space. The match for the orthorhombic phase is inconclusive -- partially because phase separation as such was not the main objective of the work done by Wang et al. Our detailed analysis (in Supplementary Section S2\,I) suggests that future experimental phase mapping and phase separation analyses would be beneficial especially in high-Cs and high-Br \ce{(Cs/FA)Pb(Br/I)3} regions to confirm if orthorhombic phase is observed. Our comparison also motivates future experimental investigations into the dependence of the material's properties on the synthesis procedure. Understanding these effects would benefit understanding perovskite stability in both short and long-term.

Although the Sn-based system holds promise for reducing the toxicity of perovskite solar cells, experimental stability data remain limited and are currently available only for the binary alloy \ce{(Cs/FA)SnI3}. Gao et al. reported successful fabrication of alloy compositions with up to 40\% of Cs \cite{gao2018robust}, whereas Pansa-Ngat et al. did not observe phase separation at Cs concentrations below 10\% \cite{pansa-ngat2021phase}. Our results are in qualitative agreement with both studies since we predict only low concentrations of Cs to be stable (up to 30\% of Cs in cubic phase at 300 K, and a stability region moderately expanded from this for \qty{150}{\degree C}).

Our results provide further insight into the quaternary \ce{(Cs/FA)Sn(Br/I)3} alloys. While its free energy landscapes closely resemble those of \ce{(Cs/FA)Pb(Br/I)3} in both shape and magnitude, the absence of central low-energy regions indicates lower stability of Sn-based quaternary alloys. Consistently, the convex hull results (Fig. \ref{fig:fig_5}) exhibit a broader stability range for \ce{(Cs/FA)Pb(Br/I)3}, suggesting fewer options for compositional engineering in the Sn-based alloy. The only exception occurs in the low-Cs, high-Br region of the composition space in the $Pnma$ phase, where more Sn-based compounds are predicted to be stable. Experimental evidence, however, suggests that $Pnma$ is not the dominant phase for these compositions \cite{wang2023sustainable}. Aside from this exception, \ce{(Cs/FA)Sn(Br/I)3} shows no stability in regions of the composition space where \ce{(Cs/FA)Pb(Br/I)3} is unstable, which indicates that future experimental efforts to optimize \ce{(Cs/FA)Sn(Br/I)3} could focus on compositions that have already been identified as favorable for the Pb-based counterpart.

Based on the curvature analysis (Figure \ref{fig:fig_5}), at room temperature (\qty{300}{K}), the highest stability regions are located near the corners of the composition space, and no additional stability is gained in the center. This limits the extent to which the quaternary perovskite alloys could be stabilized through mixing entropy, which is at its highest in the middle of the composition space. Regarding the temperature dependence of our results, we observe the same overall trend at a typical perovskite synthesis temperature of \qty{150}{\degree C}, as shown in Supplementary Fig.~S14.

In this work, the stability of the perovskite alloys was investigated in a specified scope that has implications on both comparing the experimental results to ours and on guiding future experiments. Our analysis focused on the intrinsic stability of perovskites against phase separation, without directly addressing decomposition mechanisms driven by external stresses such as moisture or air exposure. Furthermore, we computed the free energies of mixing within each phase separately, which restricts the analysis to cases of phase separation where the decomposition products retain the same lattice type as the original compound.

Direct comparison of structural phases in perovskites requires accounting for vibrational entropy contributions, as neglecting them leads to a consistent preference for the lowest-symmetry $Pnma$ phase due to its lower internal energy. Including vibrational entropy would also enable meaningful comparisons with non-perovskite phases, such as the optoelectronically inactive $\delta$-phase, offering deeper insight into decomposition pathways from active to inactive materials. While ML offers a promising route for efficiently estimating vibrational entropies through accelerated phonon calculations, the interatomic potentials developed in this work were not sufficiently accurate for that purpose. We are currently working to enhance our workflow to support efficient training with higher-tier DFT calculations, thereby enabling reliable phase comparisons.

In conclusion, we have developed an efficient computational workflow to predict the phase stability of the quaternary hybrid perovskite alloys \ce{(Cs/FA)Pb(Br/I)3} and \ce{(Cs/FA)Sn(Br/I)3}. Our results reveal that the Pb-based alloy is stable accross a broader compositional range than its Sn-based counterpart, suggesting fewer opportunities for materials optimization through compositional engineering in \ce{(Cs/FA)Sn(Br/I)3}. Furthermore, the most stable compositions are observed at low substitution fractions, indicating that mixing entropy does not stabilize these alloys in the middle of the composition range. Our approach provides a general framework for accurate and computationally efficient finite-temperature modeling of structurally complex alloys and can be extended to other mixed hybrid perovskites, accelerating the discovery and optimization of new solar cell materials.

\section{Methods}
\subsection{DFT calculations}
The DFT calculations for the MACE model data generation were carried out using the all-electron code FHI-aims \cite{blum2009ab, havu2009efficient, Xinguo/implem_full_author_list,levchenko2015hybrid}, with the Perdew–Burke–Ernzerhof (PBE) exchange-correlation functional \cite{perdew1996generalized}. To account for van der Waals interactions, we employed a non-local many-body dispersion correction \cite{hermann2020density}. Brillouin-zone integrations were performed using 4$\times$4$\times$4 and 8$\times$8$\times$8 $\Gamma$-centered $k$-point meshes for the 2$\times$2$\times$2 and 1$\times$1$\times$1 perovskite supercells, respectively. All calculations used the standard tier-2 basis sets and "tight" integration grids provided by FHI-aims. Scalar relativistic effects were treated using the zeroth-order regular approximation \cite{lenthe1993relativistic}.

\subsection{MACE interatomic potentials}
We employ MACE, an equivariant graph neural network model \cite{batatia2022design, batatia2022mace}, to fit ML interatomic potentials for structural relaxation of \ce{(Cs/FA)Pb(Br/I)3} and \ce{(Cs/FA)Sn(Br/I)3} hybrid perovskite structures. The DFT training data for the models were generated through a data generation workflow that we developed in our previous work \cite{homm2025efficient}. This workflow utilizes clustering and active learning for diverse and efficient sampling of the structural space. For this study, only minor modifications were made to the workflow: treatment of orientable molecules was implemented to account for the FA cations, and an uncertainty estimation approach utilizing an ensemble of ML models was introduced into the active learning process to enable the transition from our previous regression model to MACE. See Supplementary Section S1\,A for a detailed description of the entire training process.

To preserve the intended lattice type during structural relaxations we imposed phase-specific constraints on the lattice parameters and the halide positions. For the cubic $Pm\overline{3}m$ phase, we forced the simulation cell to maintain a cubic shape by allowing only isotropic scaling, and prohibited any tilting of the \ce{$BX$6} coordination octahedra. The tetragonal $P4/mbm$ and $I4/mcm$ phases were treated with a different set of constraints: octahedral tilting was permitted only around the $c$-axis, and the cell was allowed to deform only by adjusting the height-to-width ratio. No constraints were applied to the orthorhombic $Pnma$ phase. Structure relaxations with both DFT and MACE were performed using the ASE package \cite{larsen2017atomic}, employing the Broyden–Fletcher–Goldfarb–Shanno (BFGS) minimizer \cite{fletcher2000practical}. The relaxations had a convergence criterion of \qty{5}{meV/\AA} on the maximum atomic force.

\subsection{Direct relaxation models}
Due to the high complexity of the \ce{(Cs/FA)$B$(Br/I)3} configuration space, the MACE-based structure relaxations are not fast enough for free energy computations. To overcome this, we accelerated the energy predictions further by using the MACE potentials to generate relaxation data for training a secondary set of ML models that predict relaxed energies directly from the unrelaxed structures -- bypassing the need for explicit relaxations during the free energy sampling. We carried out this structure-to-energy mapping employing an ML model architecture that first constructs a global structural representation of an atomic geometry based on the neural network node features extracted from the the fitted MACE models. The structural representation is then mapped to an energy value using Kernel Ridge regression (KRR). 

The data generation strategy for these models was based on the same clustering and active learning workflow used for the MACE potentials, with two key modifications. First, during initial data generation, MC sampling was introduced alongside clustering to explore low-energy structures more efficiently. Secondly, the acquisition strategy of the active learning stage was modified to prioritize low-energy alloy configurations over structures with high prediction uncertainty. A detailed description of the direct relaxation model architecture and model training is available in Supplementary Section S1\,B.

\subsection{Stability analysis via free energy calculations}
Using the direct relaxation models, we were able to analyze the thermodynamic stability across the composition spaces via computing the free energies and their curvatures. To access the free energies of the perovskite alloys, we employed the Wang-Landau algorithm (see Supplementary Section S1\,C for details on our implementation), which is an MC method for sampling the temperature-independent density of states (DOS) of a system \cite{wang2001efficient}.

The algorithm estimates the DOS with a discrete grid function, $\rho(E),$ which allows approximating the partition function by summing over the energy grid:
\begin{align}
    Z \approx \sum_{i}\rho(E_i)e^{-E_i/k_BT},
\end{align}
where $T$ is the simulated temperature. The Helmholtz free energy can then be computed as:
\begin{align}
    F = -k_BT\ln Z.
\end{align}
To facilitate interpretation, we calculated the Helmholtz free energy of mixing
\begin{align*}
    \Delta F_{\text{mix}} =& F_{\ce{(Cs/FA)$B$(Br/I)3}} - xyF_{\ce{Cs$B$Br3}}\\
    &- x(1-y)F_{\ce{Cs$B$I3}} - (1-x)yF_{\ce{FA$B$Br3}}\\
    &- (1-x)(1-y)F_{\ce{FA$B$I3}}, \numberthis
\end{align*}
where $x$ and $y$ denote the elemental concentrations in \ce{Cs_xFA_{1-x}$B$(Br_yI_{1-y})3}. We performed the Wang-Landau simulation at every allowed alloy composition within the 2$\times$2$\times$2 supercell to construct the free energy landscapes.

The local stability of an alloy composition against phase separation is determined by the curvature of the free energy surface at that point. We estimated the curvatures from the simulated free energy values by first fitting a cubic spline $f(x,y)$ on the data, and then solving the eigenvalues of the Hessian matrix
\begin{align}
\matr{H}_f(x, y) &= 
\begin{bmatrix}
\frac{\partial^2 f}{\partial x^2} & \frac{\partial^2 f}{\partial x \partial y} \\
\frac{\partial^2 f}{\partial y \partial x} & \frac{\partial^2 f}{\partial y^2}
\end{bmatrix}.
\end{align}
The smaller of the two eigen values of $H_f$ corresponds to the minimum normal curvature of the free energy surface at concentration ($x$, $y$), reflecting the energetic tendency of phase separation along the most favorable direction. A positive curvature indicates local convexity of the free energy surface, and thus local stability against infinitesimal perturbations.

In addition to local stability, we assessed global thermodynamic stability by determining the convex hulls of the free energy surfaces using the spline fits. A composition lying on the convex hull is thermodynamically stable against phase separation into any other configuration of the same phase. The convex hulls for the binary alloys were determined separately by only considering decomposition along a single alloy dimension. Thus, they can suggest broader stability regions than those indicated by the full two-dimensional convex hull analysis.

\section{Data availability}
We made the DFT data generated for MACE model fitting and testing available on NOMAD (https://doi.org/10.17172/NOMAD/2026.03.17-1). The same data in processed form as well as final fitted MACE models for both alloy systems are available through Zenodo (https://doi.org/10.5281/zenodo.18957344).

\section{Code availability}
The codes used for all computational steps have been uploaded to GitLab \linebreak (https://gitlab.com/cest-group/learnsolar-hybrids).

\begin{acknowledgments}
The authors wish to acknowledge Pascal Henkel, Jingrui Li, and Miika Rasola for insightful discussions. This study was supported by the Academy of Finland through Project No.~334532. A.T. acknowledges funding from the European Union’s Horizon 2020 research and innovation programme under the Marie Sklodowska-Curie grant agreement No.~010598. We further acknowledge CSC -- IT Center for Science, Finland the Aalto Science-IT project for generous computational resources.
\end{acknowledgments}


\bibliography{ref}

\end{document}


\title{Supplemental Material: \\
Charting the thermodynamic stability of hybrid perovskite alloys with machine learning}

\author{Jarno Laakso}
\affiliation{%
 Department of Applied Physics, Aalto University, Espoo, Finland
}%

\author{Armi Tiihonen}
\affiliation{%
 Department of Mechanical Engineering, Chalmers University of Technology, Gothenburg, Sweden
} 

\author{Patrick Rinke}
\affiliation{%
 Department of Applied Physics, Aalto University, Espoo, Finland
}%
\affiliation{Physics Department, Technical University of Munich, Garching, Germany}
\affiliation{Atomistic Modelling Center, Munich Data Science Institute, Technical University of Munich, Garching, Germany}
\affiliation{Munich Center for Machine Learning (MCML)}

\date{\today}

\maketitle
\tableofcontents

\clearpage
\section{Methodology}

\subsection{\label{ssec:mace}MACE interatomic potential for faster relaxations}

\begin{figure}
    \includegraphics[]{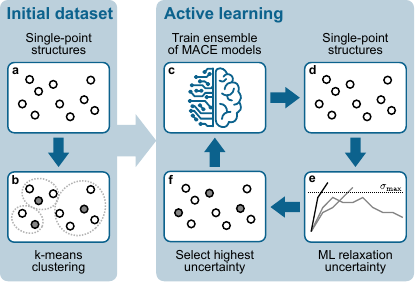}
    \caption{
    The computational workflow for training the MACE potentials. First, an initial dataset of atomic structures is constructed by sampling data for DFT labeling from a randomly generated data pool via k-means clustering. Then, the relaxation accuracy of the MACE models is improved iteratively through active learning.
    }
    \label{fig:fig_workflow_mace}
\end{figure}

We utilize MACE, an equivariant graph neural network model \cite{batatia2022design, batatia2022mace} to fit an ML interatomic potential for structural relaxation of \ce{(Cs/FA)$B$(Br/I)3} hybrid perovskite structures. The DFT training data for the models were generated through a data generation workflow that we developed for an earlier work \cite{homm2025efficient}. Here, we present the modifications to that workflow to account for the increased structural complexity of the hybrid perovskite alloys.

\subsubsection{Model training\label{ssec:mace_training}}
The DFT data generation workflow, shown in Fig. \ref{fig:fig_workflow_mace}, consists of two main steps, where clustering is first used to select a diverse set of atomistic structures for DFT labeling, and then the model accuracy is improved iteratively through active learning. To enable the modeling of various perovskite phases, we incorporated structures from four distinct lattice types -- $Pm\overline{3}m$, $P4/mbm$, $I4/mcm$, and $Pnma$ -- at all stages of the data generation process. We applied the same process separately to both \ce{(Cs/FA)Sn(Br/I)3} and \ce{(Cs/FA)Pb(Br/I)3}.

\begin{figure}
    \centering
    \includegraphics[]{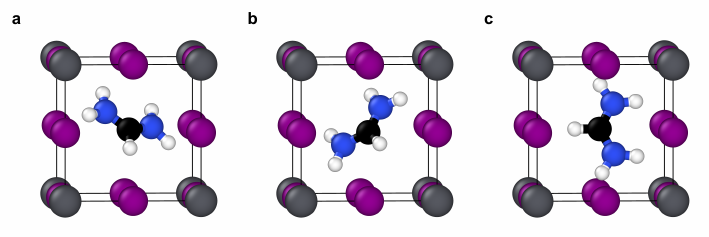}
    \caption{
    FA orientations. (a), (b), and (c) depict the three unique local minimum FA orientations obtained from the local optimum search performed with $1\times 1\times 1$ unit cells, containing $B$-site atoms (gray), $X$-site atoms (purple), carbon (black), nitrogen (blue), and hydrogen (white).
    }
    \label{fig:fig_fa_orientations}
\end{figure}

A large pool of candidate atomistic structures was required to select a diverse initial dataset via clustering. To generate alloy structures, we first needed to determine approximate geometries for the four corners of the composition space: \ce{Cs$B$Br3}, \ce{Cs$B$I3}, \ce{FA$B$Br3}, and \ce{FA$B$I3}. For the inorganic compounds \ce{Cs$BX$3} this was straightforward -- we simply performed a local minimization of the geometries using DFT with the $2\times 2\times 2$ supercell. For the organic perovskites \ce{FA$BX$3}, obtaining the global optimum geometries within a $2\times 2\times 2$ supercell is not tractable with DFT due to the vast number of possible FA orientations. Therefore, we opted to perform a configuration search over possible FA orientations within the $1\times 1\times 1$ unit cell, instead. The search identified three unique local minimum FA orientations, depicted in Fig. \ref{fig:fig_fa_orientations}. Since the perovskite phases other than the cubic $Pm\overline{3}m$ can not be represented within this minimal unit cell, we assumed the same \ce{$BX$6} octahedral tilting angles for \ce{FA$B$Br3} and \ce{FA$B$Cl3} as those found for their corresponding inorganic counterparts, \ce{Cs$B$Br3} and \ce{Cs$B$Cl3}. We then scaled the simulation cells to match the density of the global minimum $1\times 1\times 1$ structures to obtain the $2\times 2\times 2$ supercell structures for \ce{FA$B$Br3} and \ce{FA$B$Cl3}. We emphasize that these geometries served only as rough approximations of the real global optimum structures. They were used as an initial guess, and the random deviations introduced in the next phase of the data generation guaranteed that the trained ML model was able to find a more accurate geometries later on.

Next, we generated a data pool of \num{90000} different $2\times 2\times 2$ supercell alloy structures (Fig. \ref{fig:fig_workflow_mace}a). All four phases were equally represented in the generated data. The \ce{Cs/FA} and \ce{Br/I} configurations as well as the orientations of the FA molecules were fully randomized. For each alloy composition, initial geometries were constructed using Vegard’s law, by linearly interpolating both lattice parameters and atomic coordinates between the pure compounds. To introduce structural diversity, we varied the lattice parameters by randomly deviating the cell volumes, height-to-width ratios (except for the cubic $Pm\overline{3}m$ phase), and the angles between lattice vectors $\vect{a}$ and $\vect{b}$ in the $Pnma$ phase. Similarly, atomic positions were perturbed by adding random deviations to the tilting angles of the $B$-site coordination octahedra, as well as all atomic site coordinates, including the FA positions. Finally, smaller deviations were introduced within each FA molecule to vary its internal geometry. Using this procedure, we generated 100 unique alloy structures per composition and phase to form the data pool. To generate test data, we repeated the process, generating one additional structure per composition and phase. From this data, we randomly selected 50 structures per phase, yielding a test set of 200 atomic structures.

Next, we calculated vector representations of the generated structures using the Many-body Tensor Representation (MBTR) \cite{huo2022unified}, employing the implementation of the DScribe library \cite{himanen2020dscribe, laakso2023updates}. Following our previous work, we included only the interatomic distance contributions and opted to omit the higher-order terms of MBTR to reduce the computational cost. We then clustered the resulting MBTR vectors into 200 clusters (Fig. \ref{fig:fig_workflow_mace}b) using the constrained k-means algorithm \cite{bradley2000constrained, levy-kramer2018k-means-constrained}, setting the minimum cluster size to 20. To sample a diverse set of structures based on the clustering results, we selected structures in batches, each batch consisting of one randomly chosen structure from every cluster. For every batch, we performed DFT calculations to obtain energies, forces, and stresses, and fit a new MACE model on the expanded training set. Model convergence was tracked by monitoring prediction errors on the test set.

We augmented the training sets iteratively through active learning (Active learning in Fig. \ref{fig:fig_workflow_mace}), following the approach detailed by Homm et al. \cite{homm2025efficient}. In each active learning iteration, we first trained an ensemble of three MACE models with different random seeds on the current dataset (Fig. \ref{fig:fig_workflow_mace}c). Then we generated one structure per concentration and phase, resulting in 225 structures per phase in total (Fig. \ref{fig:fig_workflow_mace}d). The structures were generated using the same procedure as for the initial data pool, with two key differences: 1) the possible orientations of the FA molecules were limited to the local minimum orientations (see Fig. \ref{fig:fig_fa_orientations}) and 2) no random deviations were added to the site positions or lattice parameters. Instead, we constructed the geometries using the exact linearly interpolated values according to Vegard's law. The generated structures were then relaxed (Fig. \ref{fig:fig_workflow_mace}e) using the MACE ensemble. Relaxations were terminated either upon convergence (\qty{5}{meV/\AA} convergence criterion) or when the standard deviation of the ensemble energy predictions exceeded a threshold of \qty{2.0}{meV/\formulaunit}. We selected the 25 structures per phase that reached the threshold in fewest iterations (Fig. \ref{fig:fig_workflow_mace}f) and continued the relaxation with DFT for an additional step, resulting in two DFT labeled atomic geometries per selected structure which were added to the training set. If fewer than 25 structures reached the uncertainty threshold, the structures with the highest final MACE ensemble uncertainty were selected instead. In such cases, DFT relaxations were initiated from the final geometry reached with the MACE model.

The convergence of the active learning model was assessed through relaxations on a test set of 100 randomly generated alloy structures. The initial structures for the relaxations were generated in the same way as the candidate initial structures during the active learning process, and they were then relaxed with the current MACE model. Because it was not computationally feasible to relax these same structures with DFT to allow direct comparison, another approach was adopted: we performed single-point DFT calculations on the MACE-relaxed geometries and monitored 1) the mean absolute error (MAE) of the MACE predicted relaxation energies in comparison to DFT and 2) the average DFT energy of the relaxed geometries. When the MACE model improves, both of these metrics should decrease.

\subsubsection{MACE model hyperparameters}
The MACE models employed a cutoff radius of \qty{5}{\AA} to define the local atomic environments (\texttt{r\_max = 5.0}). The environments were expanded in a basis of five radial functions (\texttt{num\_cutoff\_basis = 5}) and spherical harmonics up to quantum number $l=2$ (\texttt{max\_ell = 2}). A correlation order of two (\texttt{correlation = 2}) was used to include pairwise and angular features explicitly, while excluding higher-order many-body terms. The network architecture comprised two message-passing layers (\texttt{num\_interactions = 2}). Hidden layers were constructed using a combination of 128 scalar channels and 128 vector channels.

The MACE models were fitted on a loss function that combined contributions from energies, forces, and stresses. To facilitate training, the DFT total energies were transformed to enthalpies of mixing:
\begin{align*}
    \Delta H_{\text{mix}} =& E_{\ce{(Cs/FA)$B$(Br/I)3}}\\
    &- xyE_{\ce{Cs$B$Br3}} - x(1-y)E_{\ce{Cs$B$I3}}\\
    &- (1-x)yE_{\ce{FA$B$Br3}} - (1-x)(1-y)E_{\ce{FA$B$I3}}, \numberthis
\end{align*}
where $E_{\ce{(Cs/FA)\textit{B}(Br/I)3}}$ is the total energy of an alloy structure, $E_{\ce{\textit{ABX}_3}}$ are the total energies of the pure perovskite structures ($E_{\ce{FA\textit{BX}_3}}$ were derived from the minimum $1\times 1\times 1$ unit cell energies), where $x$ and $y$ denote the elemental concentrations in \ce{Cs_xFA_{1-x}$B$(Br_yI_{1-y})3}. The number of epochs scaled with the number of structures ($N_{\text{train}}$) in the training set through
\begin{align}
    N_{\text{epochs}} = \frac{10^6}{N_{\text{train}}}.
\end{align}
For the final 20\% of the epochs, the weight of the energy term in the loss function was increased by a factor of 1000. Optimization was performed using the AMSGrad algorithm with exponential moving average (\texttt{ema\_decay = 0.99}).

\subsection{\label{ssec:drm}Direct relaxation model}
Due to the high complexity of the \ce{(Cs/FA)$B$(Br/I)3} configuration space, the MACE-based structure relaxations are not fast enough for free energy computations. To overcome this, we accelerated the energy predictions further by utilizing the MACE potentials to generate training data for a secondary set of eight phase- and material-specific ML models that predict relaxed energies directly from the unrelaxed structures, thereby bypassing the need for explicit relaxations during the free energy sampling. This approach could not have been adopted without first training the MACE model because generating sufficient relaxation data directly with DFT would have been prohibitively expensive. We carried out this structure-to-energy mapping employing an ML model architecture, where structural representations obtained from the fitted MACE models were combined with Kernel Ridge regression (KRR). Our data generation approach here was similar to that used for the MACE potentials, incorporating both clustering and active learning. However, changes detailed later in this section were implemented to the active learning acquisition strategy to achieve better performance. Since the relaxation constraints were phase-specific, a separate model was fitted for each phase. Furthermore, we introduced a completely new stage to the data generation workflow, utilizing Monte Carlo (MC) sampling to explore low-energy structures more efficiently.

\subsubsection{ML methodology}
We initially attempted using the standard MACE model for learning the mapping between unrelaxed structures and relaxed energies. However, this approach yielded suboptimal results, likely due to the fact that the learning task is now inherently non-differentiable and thus atomic forces and stresses could not be utilized in the model fitting. Moreover, directly predicting relaxed energies from unrelaxed structures may conflict with the assumption of locality that underpins the interatomic potentials such as MACE. To address these problems, we instead implemented an ML model that utilizes global structure descriptions derived from the the structure embeddings of the trained MACE potentials.

MACE computes feature vectors for each atom in a geometry, which are iteratively updated during message passing iterations based on the local atomic environment. To construct a global representation out of these atomic features, we extracted the invariant node features from all message passing layers of the fitted MACE models and aggregated them based on element. Specifically, the $i$th component of an aggregated vector corresponding to element A was given by:
\begin{align}
    \left(\vect{x}^{A}\right)_i = \sum_{j\in A}D_{ji},
\end{align}
where $\matr{D}$ is a matrix that contains all invariant features, with each row corresponding to an atom. We then concatenated the elemental contributions together to form the final representation vectors $\vect{x}$. For structures lacking a particular element, the corresponding section in the representation was filled with zeros to ensure consistent dimensionality across all features. The dimensionality of the final vectors depends on the hyperparameters of the MACE model and was \num{1792} in our case.

We mapped the global representation vectors to the corresponding MACE-relaxed energies $E_{\text{relax}}^{\text{MACE}}$ using KRR. The predicted energy for a structure represented by feature vector $\vect{x}$ is given by
\begin{align}
    E_{\text{relax}}^{\text{KRR}} = \sum_{i}^{N}\beta_i k(\vect{x},\vect{x}_i)\,,
\end{align}
where $\beta_i$ are regression coefficients, $k$ is a kernel function, and $\vect{x}_i$ are a set of $N$ training features. We used the Gaussian kernel function
\begin{align}
    k(\vect{x},\vect{x}^{\prime}) = \exp(-\gamma ||\vect{x}-\vect{x}^{\prime}||_2^2)\,,
\end{align}
where $\gamma$ is a hyperparameter that controls the width of the kernel distribution. The regression coefficients $\beta_i$ are obtained by solving
\begin{align}
    \vect{\beta} = (\matr{K}+\alpha\matr{I})^{-1}\vect{E}_{\text{relax}}^{\text{MACE}}\,,
\end{align}
where $\matr{K}$ is the kernel matrix $K_{i,j}:=k(\vect{x}_i,\vect{x}_j)$, $\vect{E}_{\text{relax}}^{\text{MACE}}$ is a vector containing the MACE-relaxed energies of a training set, and $\alpha$ is a regularization parameter.

\subsubsection{\label{ssec:drm_data_generation}Data generation}

\begin{figure}
    \includegraphics[]{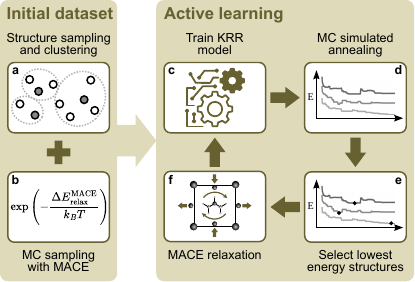}
    \caption{
    The computational workflow for training the direct relaxation models. First, an initial dataset of atomic structures is generated through a combination of sampling via clustering and MC simulations. Then, the accuracy of the model in the low-energy regime is improved iteratively with active learning.
    }
    \label{fig:fig_workflow_relax}
\end{figure}

The data generation workflow utilized for the direct relaxation models was similar to the one used for training the MACE potentials. The workflow, that is depicted in Fig. \ref{fig:fig_workflow_relax}, consisted of two main stages: 1) initial data generation through clustering and MC sampling, and 2) improving model predictions with active learning.

We began by sampling configurationally diverse data via clustering (Fig. \ref{fig:fig_workflow_relax}a). To that end, we generated a large pool of unrelaxed structures for each phase. The same structure generation scheme was used here as during the active learning stage of the MACE model training -- incorporating randomized Cs/FA and Br/I alloy configurations and FA orientations, but without applying any deviations to lattice parameters or atomic position. A total of 400 structures were generated for each of the 225 alloy concentrations, resulting in data pools with \num{90000} structures for each phase.

For each phase, the global feature vectors of the generated structures were clustered into \num{2000} clusters, from which structures were selected in batches of \num{2000} for labeling. A larger number of clusters was used here compared to the MACE training workflow to account for the slower learning rate. The energy labels were obtained by relaxing all selected structures with the MACE potential and assigning the predicted energy of the final relaxed geometry as the label. The training process was monitored by evaluating model prediction errors after each batch on eight phase-specific test sets containing \num{4500} structures each.

Lower-energy structures contribute disproportionately to the free energy, making accurate predictions in this energy range particularly important for the KRR models. However, the randomly sampled data pools contained relatively few low-energy structures, leading to poor sampling efficiency of this critical region with the clustering method. To address this problem, we introduced a new sampling method aimed specifically to incorporate more low-energy structures (Fig. \ref{fig:fig_workflow_relax}b). We employed the trained MACE potentials in MC simulations at constant temperature of \qty{300}{K}. The MC sampling biases the structure exploration toward the energy range that is the most relevant for the free energy calculations.

At each alloy concentration, we initialized the MC algorithm with randomized alloy configurations and FA orientations. In each subsequent iteration, the MC algorithm generated a new configuration by proposing one of three modifications to the structure: 1) a swap of a random pair of Br and I atoms, 2) a swap of a random pair of Cs and FA, or 3) a reorientation of a randomly selected FA molecule to another local minimum configuration. Each modification type was selected with equal probability. We then relaxed the atomic structure using the MACE model, and compared the relaxed energy to the energy before the proposed modification. Following the Metropolis principle, energetically favorable changes were always accepted, while unfavorable ones were accepted with the probability
\begin{align}
    P(\text{accept}) = \exp\left(-\frac{\Delta E_{\text{relax}}^{\text{MACE}}}{k_BT}\right)\,,
\end{align}
where $T$ is the simulated temperature. If accepted, the modified structure was used as the starting point for the next iteration.

We ran the MC algorithm for \num{2000} steps at $T=\qty{300}{K}$ for each composition. Since the MC algorithm can revisit the same structure multiple times, we removed the duplicate structures from the data. From the remaining unique structures, we selected up to 20 configurations per alloy composition, spaced uniformly across the sampled energy range. For compositions that had fewer than 20 unique structures, all available structures were included.

Although the addition of low-energy structures via MC sampling improves the models' prediction accuracy in that energy regime, the models can still predict artificially low energies for structures that are, in reality, high in energy, leading to inaccurate free energy calculations. To fix this issue, we once again utilized active learning. We found the uncertainty-based active learning approach, similar to the one used during MACE training, ineffective in this context. Instead, we developed another approach based on energy minimization via MC simulation guided by the current KRR model.

At the start of each active learning iteration, we fitted a KRR model on the current data set (\ref{fig:fig_workflow_relax}c). This model was then used to search for the global minimum energy configuration for each of the 225 alloy compositions via MC simulated annealing simulations (\ref{fig:fig_workflow_relax}d). The MC algorithm followed the same sampling scheme used in the earlier MC-based data generation stage, with two key modifications: the current KRR model was now used for the prediction of relaxed energies instead of MACE, and the simulated temperature of the system was linearly decreased during the simulation from \qty{300}{K} to \qty{0}{K}. After each \num{10000}-step MC run, we selected the structure with the lowest energy that had not yet been added to the training set (\ref{fig:fig_workflow_relax}e) and relaxed it with the MACE potential to generate the energy label (\ref{fig:fig_workflow_relax}f). One MC minimization was performed per composition in each active learning cycle, resulting in the addition of up to 225 new structures per iteration.

\subsection{\label{ssec:wang_landau}Wang-Landau Algorithm}
Wang-Landau algorithm is an MC method for sampling the temperature-independent density of states (DOS) of a system \cite{wang2001efficient}. Our implementation of the Wang-Landau random walk resembled the MC simulations detailed in Section \ref{ssec:drm_data_generation}, including an identical protocol for the proposal of new configurations. The algorithm estimated the DOS with a discrete grid function $\hat{\rho}(E).$ During each iteration of a random walk, the energy of the proposed state was evaluated with the direct relaxation model, and the proposed configuration change was accepted with probability
\begin{align}
    P(\text{accept}) = \min\left(1, \frac{\hat{\rho}(E_{\text{current}})}{\hat{\rho}(E_{\text{proposed}})}\right),
\end{align}
where $E_{\text{current}}$ and $E_{\text{proposed}}$ were the energies of the current and proposed configurations, respectively. After every iteration, $\hat{\rho}(E)$ was updated based on the energy of the current state according to 
\begin{align}
    \hat{\rho}(E_{\text{current}}) \leftarrow f\hat{\rho}(E_{\text{current}}),
\end{align}
where $f$ is a scalar update factor that decreases over the course of the simulation. The value of $f$ was decreased based on the histogram $H(E)$, which tracked the number of times each energy bin had been visited by the algorithm since the previous update of $f$. In our case, when the value of the least-sampled bin was at least 50\% of the histogram mean, we updated $f$:
\begin{align}
    f \leftarrow \sqrt{f}
\end{align}
However, we always ran the simulation for at least \num{10000} iterations after updating $f,$ regardless of the flatness of $H.$ We initialized $f$ with the value $e$ and terminated the simulation once $f$ dropped below $\exp(10^{-3})$.

We performed the Wang-Landau simulation at every alloy composition. The resulting $\hat{\rho}(E),$ however, required normalization before they could be used for calculating free energy values. We used the total number of configurations as a normalization factor:
\begin{align}
    N_{\text{config}} = \frac{N_A!}{N_{\ce{Cs}}!N_{\ce{FA}}!} \times \frac{N_X!}{N_{\ce{Br}}!N_{\ce{I}}!} \times 72^{N_{\ce{FA}}},
\end{align}
where the first term is the number of $A$-site permutations, second term is the number of $X$-site permutations, and the third term is the number of FA orientation combinations. The normalized DOS was then given by
\begin{align}
    \rho(E) = N_{\text{config}}\frac{\hat{\rho}(E)}{\sum_i \hat{\rho}(E_i)}.
\end{align}

The choice of 72 as the number of possible orientations for a single FA molecule in the definition of $N_{\text{config}}$ was based on the number of local minima identified for the FA molecule within the $1 \times 1 \times 1$ supercell during the configuration search described in Section \ref{ssec:mace_training}. This value may therefore not exactly reflect the actual number of accessible FA orientations in a $2 \times 2 \times 2$ supercell, when structural relaxations are incorporated. However, a change in this number corresponds to a linear transformation in free energy with respect to the alloying concentrations. Therefore, the specific choice of 72 does not affect the outcome, since only the free energy of mixing with respect to the the pure perovskite compounds was considered.

\newpage
\section{Results}
\subsection{\label{ssec:mace_initial}Convergence tests for MACE potentials trained on initial clustering-based data}
\begin{figure}[h]
    \includegraphics[]{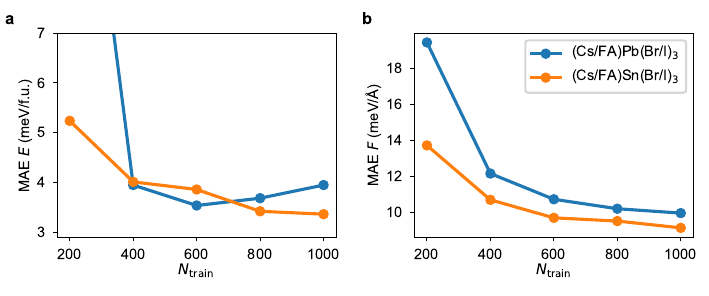}
    \caption{%
    Convergence test results for the MACE models during the initial data generation stage via clustering for the lead- (blue) and tin- (orange) based perovskites. Single-point (a) energy and (b) force prediction learning curves (mean absolute error of predictions on the test dataset as a function of the number of training structures), respectively, during the initial dataset generation.
    }
    \label{fig:fig_mace_sp_lc}
\end{figure}

During the generation of the initial datasets for the MACE models, the DFT calculations were added to the training sets in batches of 200. After each batch, the model prediction errors were evaluated on the test sets comprising DFT single-point calculations of 200 alloy structures (see Section \ref{ssec:mace} for details on test data generation). For both \ce{(Cs/FA)Pb(Br/I)3} and \ce{(Cs/FA)Sn(Br/I)3}, the mean absolute errors (MAE) of the energy and force predictions first decrease rapidly after every added batch (Fig. \ref{fig:fig_mace_sp_lc}a and \ref{fig:fig_mace_sp_lc}b). After five batches the errors have effectively converged, which is why we decided to stop adding data at that point.

\subsection{\label{ssec:mace_al}Convergence tests for active learning of MACE potentials}
\begin{figure}[h]
    \includegraphics[]{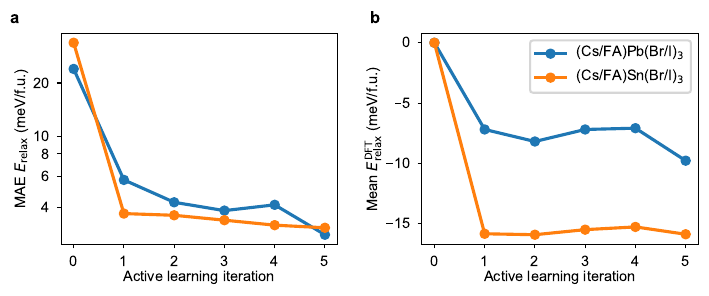}
    \caption{%
    Convergence test results for the MACE models during active learning for the lead- (blue) and tin- (orange) based perovskites. 
    (a) Mean absolute errors of the MACE energy predictions on the relaxed test structures in comparison to DFT.
    (b) Mean DFT energies of the MACE-relaxed test set structures during active learning. The values have been shifted so that the mean energies at iteration 0 (before active learning) are zero.
    }
    \label{fig:fig_mace_al_lc}
\end{figure}

The relaxation accuracy of the MACE potentials was enhanced through active learning. After each active learning step, we used the current fitted models to relax the test set of 100 alloy structures and performed DFT single-point calculations on the resulting geometries (see Section \ref{ssec:mace} for details). The learning curves that depict the prediction errors on these relaxed test geometries is shown in Fig. \ref{fig:fig_mace_al_lc}a. The errors decrease initially, but after a few iterations they saturate.

To further judge the convergence of the MACE potentials, we analyzed the DFT single-point energies of the MACE-relaxed structures. Figure \ref{fig:fig_mace_al_lc}b shows the mean DFT energies for the two potentials over the active learning process. Again, the mean energies decrease rapidly initially, showing that the models are improving and finding lower-energy configurations. After about three iterations, however, the mean energies plateau, suggesting convergence.

Based on these tests, we stopped the active learning after five iterations. The final training sets consisted of \num{2000} atomic structures, representing an order of magnitude reduction compared to the ML model utilized in our previous studies \cite{laakso2022compositional, homm2025efficient}. Given the fast learning rate, even lower prediction errors could probably be achieved by increasing the MACE model complexity and expanding the training data, if needed in future applications.

\newpage
\subsection{Analysis of initial dataset generation for direct relaxation models via clustering}\label{ssec:init_direct}
\begin{figure}[h]
    \includegraphics[]{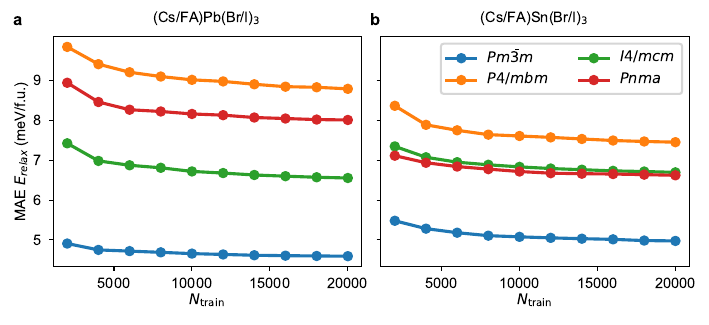}
    \caption{%
    Convergence test results for the direct relaxation models during initial data generation via clustering. Mean absolute errors of predictions in comparison to MACE relaxation energies as a function of training set size for the four phases of (a) \ce{(Cs/FA)Pb(Br/I)3} and (b) \ce{(Cs/FA)Sn(Br/I)3}.
    }
    \label{fig:fig_relax_clustering}
\end{figure}

The initial data generation for the direct relaxation models was done by clustering large pools of MACE relaxation data into \num{2000} clusters, as described in Section \ref{ssec:drm_data_generation}. Structures were selected for labeling based on the clustering results in batches of \num{2000}, and after each batch, the current model errors were evaluated on the test sets. The resulting learning curves for all eight phase- and material-specific direct relaxation models are presented in Fig. \ref{fig:fig_relax_clustering}. Notably, the reduction in errors is minimal after the first batch, with less than \qty{1}{meV/\formulaunit} improvement for any model throughout the data generation process. This convergence in accuracy motivated our decision to only include the first three batches from the clustering in the initial datasets.

\newpage
\subsection{Hyperparameter optimization for direct relaxation models}
The hyperparameters of the KRR model used in the direct relaxation models were optimized based on the initial datasets before initiating active learning. The model has two hyperparameters: regularization parameter $\alpha$ and the parameter that controls the kernel function width $\gamma.$ We optimized these hyperparameters with grid search using a $11\times 9$ grid spanning from $\alpha=10^{-10}$ to $\alpha=10^{-5}$ and from $\gamma=10^{-7}$ to $\gamma=10^{-3}.$ For each set of hyperparameter values, we evaluated the model prediction errors of the four phase-specific models for both \ce{(Cs/FA)Pb(Br/I)3} and \ce{(Cs/FA)Sn(Br/I)3} utilizing 5-fold cross-validation. The hyperparameter optimization was performed before active learning on the datasets consisting of data from both clustering and MC sampling.

\begin{figure}[h]
    \includegraphics[]{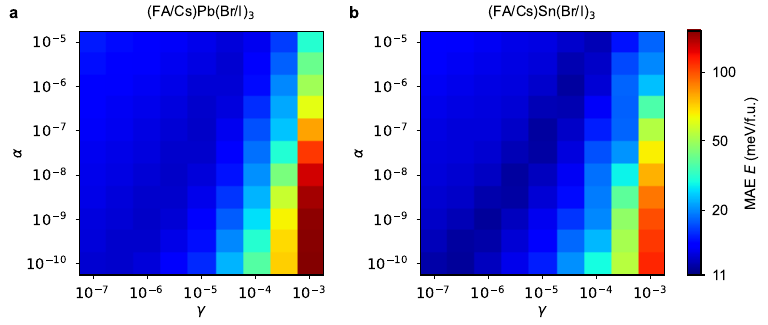}
    \caption{%
    Grid search results from the optimization of direct relaxation model hyperparameters $\alpha$ and $\gamma$ for (a) \ce{(Cs/FA)Pb(Br/I)3} and (b) \ce{(Cs/FA)Sn(Br/I)3}. The results shown are an average of the cross validated mean absolute errors over the four phases.
    }
    \label{fig:fig_hp}
\end{figure}

The hyperparameter optimization results obtained by averaging the cross-validated mean absolute errors (MAEs) over the four phases are shown in Fig. \ref{fig:fig_hp}. Slightly different minima are obtained for the two alloy systems, but $\alpha=10^{-7}$ and $\gamma=10^{-5}$ result in the lowest average error over the two. The same hyperparameter values were used for all the models to ensure consistency between the four phases.

\newpage
\subsection{Active learning of direct relaxation models}\label{ssec:al_relax}
\begin{figure}[h]
    \centering
    \includegraphics{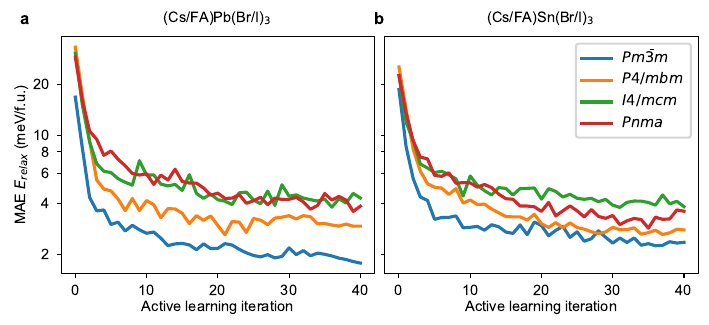}
    \caption{
    Convergence test results from the active learning stage of the direct relaxation model training for the (a) lead- and (b) tin-based perovskites across four phases. At each active learning iteration, the current models were employed to identify the minimum-energy configuration for every alloy compositions. The mean absolute errors of the predicted energies of these configurations in comparison to MACE-relaxed energies are shown.
    }
    \label{fig:fig_relax_al_lc}
\end{figure}

The final training step for the direct relaxation models focused on improving their predictions in the low-energy regime using active learning. To monitor this process, we evaluated the direct relaxation model prediction error during each iteration on the selected structures -- obtained through energy minimization -- before adding these structures to the training set. Since we ensured that no duplicate structures were included, these error evaluations remained unbiased. Figure \ref{fig:fig_relax_al_lc} shows the MAEs across active learning iterations for all the phases. All models improve quickly with added data. The average errors after 40 active learning iterations are \qty{3.2}{meV/\formulaunit} and \qty{3.1}{meV/\formulaunit} for \ce{(Cs/FA)Pb(Br/I)3} and \ce{(Cs/FA)Sn(Br/I)3}, respectively.

\begin{figure}[h]
    \includegraphics[]{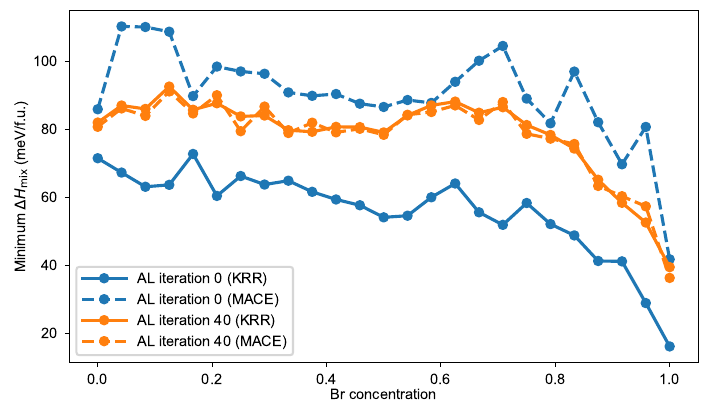}
    \caption{%
    Comparison of the energy minimization performance of the direct relaxation models for the $Pm\overline{3}m$ phase of \ce{(Cs/FA)Pb(Br/I)3} at active learning iterations 0 (blue) and 40 (orange). The KRR model's predictions for the minimum-energy configurations with zero Cs concentration are shown with solid lines and the corresponding MACE relaxation energies with dashed lines.
    }
    \label{fig:fig_relax_al_minima}
\end{figure}

At the beginning of active learning, the prediction errors for the minimum-energy structures obtained with the direct relaxation models are very high, averaging approximately \qty{20}{meV/\formulaunit}. To better understand the source of these large errors, we analyzed the results of the energy minimization runs performed during the active learning and once after it for the final models. Figure \ref{fig:fig_relax_al_minima} illustrates a representative example case ($Pm\overline{3}m$ phase of \ce{(Cs/FA)Pb(Br/I)3} at $x_{\text{Cs}}=0$) of the direct relaxation model and MACE relaxation predictions for the minimum-energy structures obtained before and after active learning. Before active learning, the minimum energies identified through MC simulations are very low, but subsequent MACE relaxations reveal that the direct relaxation model severely underestimates the true energies. By the final active learning iteration, the obtained minimum energies are higher but much closer to the corresponding MACE relaxation energies. Moreover, when comparing the the MACE relaxation energies of the minimum-energy structures obtained during the first and final active learning iterations, the energies in the final iteration are the same or lower than those in the first one.

The analysis indicates that the high errors in the initial iterations of active learning arise from a combination of two issues: 1) generally poor accuracy in the low-energy region, and 2) even larger errors for individual higher-energy structures, which leads to the minimization algorithm incorrectly identifying them as energy minima. With active learning, both of these issues were substantially mitigated, leading to to significantly improved accuracy in energy minimization.

\newpage
\subsection{Free energy data cleaning} \label{ssec:energy_data_cleaning}
\begin{figure}[h]
    \includegraphics[]{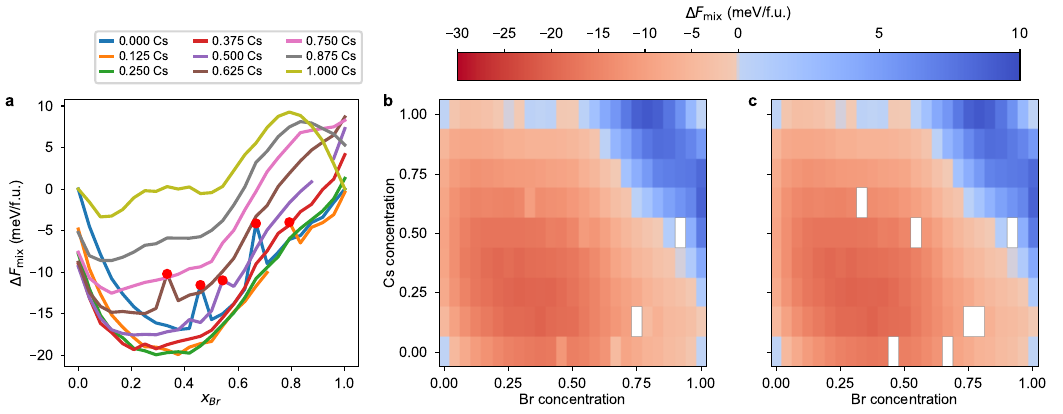}
    \caption{
    Free energy data cleaning for the $Pm\overline{3}m$ phase of \ce{(Cs/FA)Pb(Br/I)3}.
    (a) Raw free energy curves ($\Delta F_\text{mix}$) for individual Cs concentrations, with the outliers deviating from the general trend marked with red circles.\linebreak
    (b) Two-dimensional free energy landscape before outlier removal.
    (c) The same landscape after outlier removal.
    }
    \label{fig:fig_outlier}
\end{figure}

Due to the relatively high cost of energy predictions, we employed lower-than-usual iteration limits for the Wang-Landau sampling. This sometimes resulted in incomplete sampling of the energy space, leading to outliers in the raw free energy data.  An example is shown for the $Pm\overline{3}m$ phase of \ce{(Cs/FA)Pb(Br/I)3} in Fig.\ref{fig:fig_outlier}a, where the raw free energy values for each Cs concentration are presented separately. The outliers, that deviate from the general trend by multiple meV/f.u. are marked with with red dots. These outliers were removed from the data before further analysis. In addition to visual inspection, the outlier identification was facilitated by the fact that Wang-Landau sampling was repeated with two random seeds -- significant discrepancies between the two runs were strong indicators of unreliable sampling. Figures~\ref{fig:fig_outlier}b and \ref{fig:fig_outlier}c compare the two-dimensional free energy landscapes before and after outlier removal.

\newpage
\subsection{\label{ssec:i4mcm_results}$I4/mcm$ free energies}
In the study, we computed free energy landscapes for four perovskite phases: $Pm\overline{3}m,$ $P4/mbm,$ $I4/mcm,$ and $Pnma.$ The phases were differentiated in computations by applying different structural constraints during relaxation, preventing symmetry breaking in the simulation cell shape and undesirable tilting of \ce{$BX$6} coordination octahedra. For the two tetragonal phases, $P4/mbm$ and $I4/mcm$, the constraints were identical, with the only distinguishing factor being the initial geometries, featuring in-phase octahedral tilts for $P4/mbm$ and anti-phase tilts for $I4/mcm$. 

\begin{figure}[h]
    \includegraphics[]{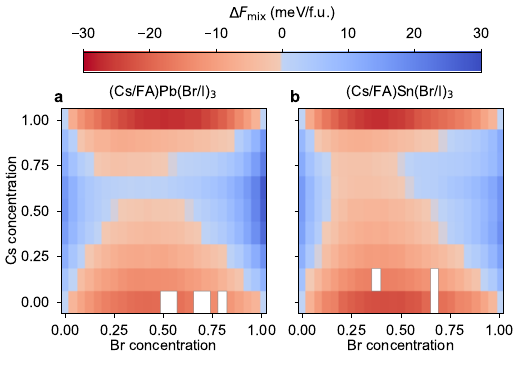}
    \caption{
    Helmholtz free energy of mixing ($\Delta F_{\text{mix}}$) landscapes at \qty{300}{K} for the tetragonal $I4/mcm$ phase of (a) \ce{(Cs/FA)Pb(Br/I)3} and (b) \ce{(Cs/FA)Sn(Br/I)3}. White rectangles imply missing data due to convergence problems of the Wang-Landau algorithm.
    }
    \label{fig:fig_i4mcm_energy}
\end{figure}

\begin{figure}[h]
    \includegraphics[]{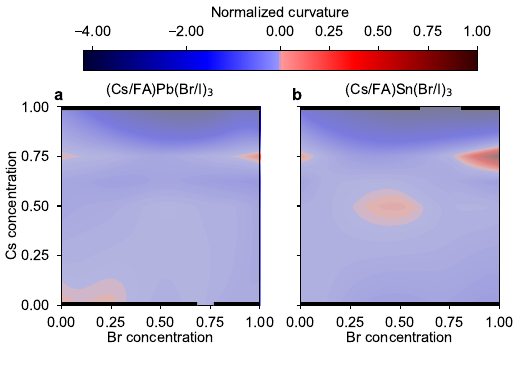}
    \caption{
    Curvature of the free energy surfaces at \qty{300}{K} for the tetragonal $I4/mcm$ phase of (a) \ce{(Cs/FA)Pb(Br/I)3} and (b) \ce{(Cs/FA)Sn(Br/I)3}. The same curvature normalization was used here as in Fig. 5 of the main text. No quaternary compositions are on the convex hull. Bold black lines on the edges indicate binary alloy compositions that are on the convex hull.
    }
    \label{fig:fig_i4mcm_curvature}
\end{figure}

However, during the data generation stage for the $I4/mcm$ direct relaxation models, we found that at some compositions the lowest-energy structures spontaneously relaxed into the $P4/mbm$ phase, even though they were initialized as $I4/mcm$. As a result, the direct relaxation models could not learn correctly the mapping between unrelaxed $I4/mcm$ structures and relaxed energies, leading to unreliable free energy results. In our current computational framework, there is no effective way to prevent this phase change during relaxation, which is why we decided to exclude the $I4/mcm$ free energy results from the main text. For completeness, we provide the computed $I4/mcm$ free energy landscapes in Fig. \ref{fig:fig_i4mcm_energy}, with the corresponding convex hulls and curvature plots shown in Fig. \ref{fig:fig_i4mcm_curvature}.

\vspace{-0.5cm}

\newpage
\subsection{Free energies at \qty{150}{\degree C}}\label{ssec:150}
\begin{figure}[h]
    \centering
    \includegraphics[width=\figurewidth]{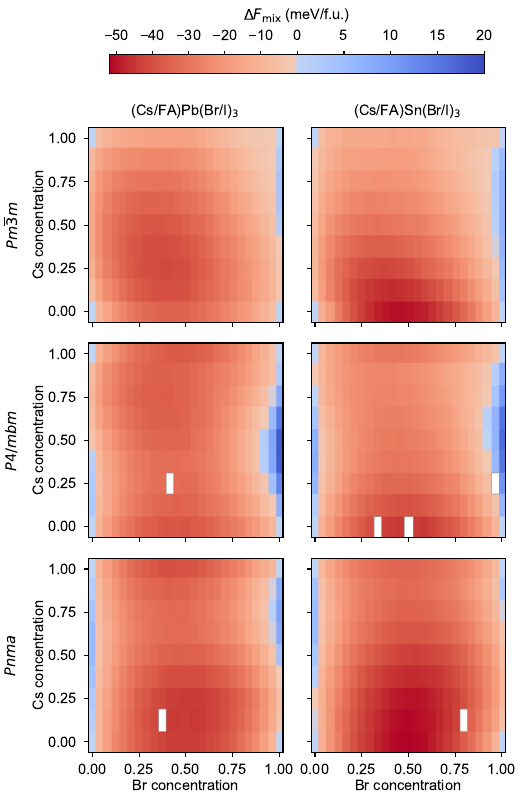}
    \caption{
    Helmholtz free energy of mixing ($\Delta F_{\text{mix}}$) landscapes at \qty{150}{\degree C} (\qty{423.15}{K}). The left column contains the results for \ce{(Cs/FA)Pb(Br/I)3} and the right column for \ce{(Cs/FA)Sn(Br/I)3}. The rows correspond to three phases: cubic $Pm\overline{3}m$, tetragonal $P4/mbm$, and orthorhombic $Pnma$. White rectangles imply missing data due to convergence problems of the Wang-Landau algorithm.
    }
    \label{fig:fig_higher_t_energy}
\end{figure}

\begin{figure}[h]
    \includegraphics[width=\figurewidth]{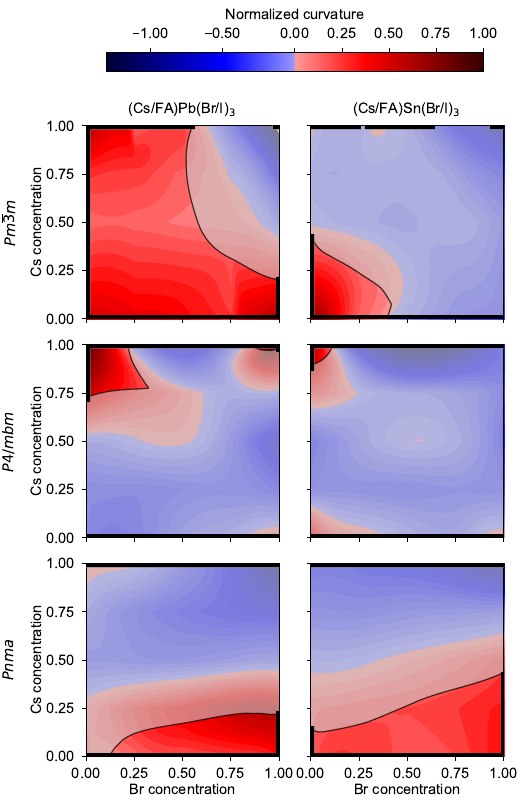}
    \caption{
    Curvature of the free energy surfaces at \qty{150}{\degree C}. The same curvature normalization was used here as in Fig. 5 of the main text. The compositions that are on the convex hull are highlighted by brighter red tones. Bold black lines on the edges indicate binary alloy compositions that are on the convex hull. The left column contains the results for \ce{(Cs/FA)Pb(Br/I)3} and the right column for \ce{(Cs/FA)Sn(Br/I)3}. The three rows correspond to the phases: cubic $Pm\overline{3}m$, tetragonal $P4/mbm$, and orthorhombic $Pnma$.
    }
    \label{fig:fig_higher_t_curvatures}
\end{figure}

In the main text, we presented our results for the free energy landscapes of \ce{(Cs/FA)Pb(Br/I)3} and \ce{(Cs/FA)Sn(Br/I)3} at \qty{300}{K} (Fig. 4 and 5). However, since perovskite solar cells are typically fabricated at much higher temperatures, it is also important to understand material behavior under those conditions. Since the Wang-Landau algorithm is temperature-independent, we are able to calculate the free energy landscapes at any temperature without repeating the computationally expensive structure sampling. The resulting free energy landscapes at \qty{150}{\degree C} are presented in Fig. \ref{fig:fig_higher_t_energy}. In comparison to the results at \qty{300}{K}, the obtained free energies are generally lower and the composition regions with positive free energies have shrunk considerably. The convex hulls and curvatures of the landscapes are shown in Fig. \ref{fig:fig_higher_t_curvatures}. While the results are similar to those at \qty{300}{K}, the stable composition regions expand with the increase in temperature.

\newpage
\subsection{Comparison of free energy results to an experimental study by Wang et al.}\label{ssec:comp}
The most comprehensive experimental study of \ce{(Cs/FA)Pb(Br/I)3} phase stability was carried out by Wang et al. \cite{wang2023sustainable}, who fabricated samples at 81 different alloy compositions with 10\% compositional steps using high-throughput synthesis equipment based on microprinting, in a laboratory space with 30\% air humidity. Thus, they succeeded in sampling this large composition space in full in their extensive and detailed work. They applied thermal processing on the samples at \qty{150}{\degree C} during synthesis and later characterized the samples via synchrotron X-ray diffraction (HT-WAXS) at room temperature in vacuum, which is why the authors state their phase diagrams should be considered non-equilibrium. The authors aimed for determining the phase and photostability maps for the compositions under these conditions. For these reasons, the experimental conditions in Wang et al.'s study do not directly match to our work that focuses on phase separation tendencies. Still, the comparison to Wang et al.'s work serves as a useful qualitative confirmation for our results.

\begin{figure}
    \includegraphics[]{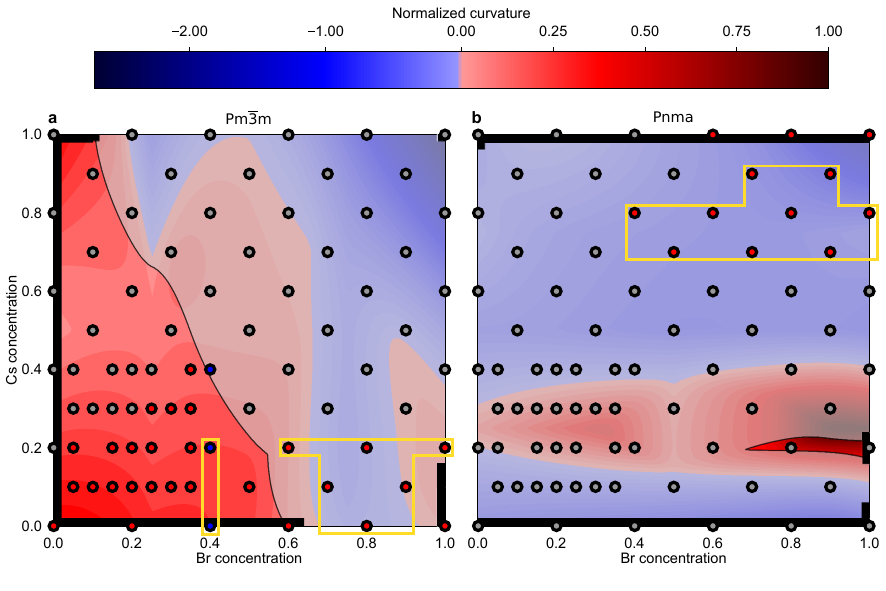}
    \caption{%
    Comparison of the convex hull and curvature results at \SI{300}{K} with experiments by Wang et al. Circles denote the experimental samples: red for single-phase samples matching the phase under consideration, blue for samples that phase-separated into two components of the same lattice type, and gray for all other cases. Composition regions where the experiments disagree with computational stability predictions are highlighted with yellow boxes.
    }
    \label{fig:fig_wang}
\end{figure}

Figure \ref{fig:fig_wang} shows the compositions examined by Wang et al., overlaid with our convex hull and curvature results at \SI{300}{K} (room temperature) for both the cubic $Pm\overline{3}m$ and orthorhombic $Pnma$ phases. The tetragonal $P4/mbm$ phase is omitted here because Wang et al. did not consider it. For both investigated phases, the experimentally tested compositions are marked as circles in three colors: red for the samples that were classified as single-phase based on HT-WAXS and matched the lattice type under consideration, blue for samples that exhibited phase separation into two components of the \textit{same} lattice type, and gray for all other cases (i.e., samples with a different lattice type, or phase separation into differing lattice types -- a case not directly comparable with our results). In our framework, stability (in the sense that the composition is on the convex hull) implies that the material should not undergo phase separation into two phases of the same symmetry. Thus, for our predictions to match perfectly with the experiments, none of the red circles should lie outside the stable regions, while no blue circles should be inside them. The apparent discrepancies are limited to three composition regions, which are highlighted with yellow boxes in Figure \ref{fig:fig_wang} and discussed below.

Our cubic phase results in Fig. \ref{fig:fig_wang}a (calculated at \qty{300}{K}) are in general in agreement with Wang et al., since ours indicate that the cubic phase ($Pm\overline{3}m$) is stable at the lower left part of the composition map (high-I and high-FA) and not stable at the high-Br side of the composition map (apart from the fully Br compositions with less than 18\% of Cs). Fig. \ref{fig:fig_higher_t_curvatures}a also illustrates that the stable regions are temperature-dependent and expand with increasing temperature. Wang et al. have also observed cubic phases (alone or together with other phases) in HT-WAXS measurements across the lower left (high-I and high-FA) triangle of the composition map, and pure cubic phases where indicated by red dots in Fig. \ref{fig:fig_wang}a. They observed the shifting of the perovskite (100) peak along with the composition change, suggesting that the phases are truly mixed.

Next, we analyze the cubic phase correspondence on a sample-to-sample basis for the two composition regions where the results are not directly matching (the yellow boxes in Figure \ref{fig:fig_wang}a). The first non-matching region is located near the \ce{FAPbBr3} corner of the composition space and consists of five samples. The seeming disagreement between Wang et al.'s experiment (single-phase cubic) and our result (phase separation) is highly likely a result from the lingering effect from the samples being synthesized at an elevated temperature and the phase map representing non-equilibrium samples as the authors also noted. Our results in Fig. \ref{fig:fig_higher_t_curvatures}a simulated at the synthesis temperature Wang et al. used indeed show that during synthesis, this region is on the convex hull. Thus, these results support ours and highlight the importance of accounting for the environment conditions the samples undergo when evaluating the stability.

The second region is located at 40\% Br and $<20$\% Cs and it contains two samples. Within the region, experiments exhibit phase separation into two $Pm\overline{3}m$ components, while our results predict single-phase stability. Sample images provided by Wang et al. in their supplementary data suggest these samples are lower quality than the most they have prepared, thus quality variations between individual samples could contribute to the disagreement. More repeated samples with investigations focused on the peak-widening would be beneficial to confirm the structure of the samples in this composition region.

Figure \ref{fig:fig_wang}b shows that we predict the orthorhombic $Pnma$ phase to be stable only for pure Cs and pure FA compositions, as well as for few high-Br compositions with approximately 20\% of Cs. In contrast, Wang et al. observed single-phase $Pnma$ samples at high-Cs (more than 70\%) high-Br (more than 40\%), which is not in agreement with our results. It is, however, noteworthy that our results for another phase, $P4/mbm$, indicate a degree of metastability around this region, especially at \qty{150}{\degree C} (see Fig. \ref{fig:fig_higher_t_curvatures}). The X-ray diffraction spectra for the $P4/mbm$ and $Pnma$ phases are very similar, which makes it challenging to identify these two phases from high-throughput-prepared samples ($P4/mbm$ was not discussed by Wang et al.). Thus, we highlight that investigations in this composition region with repeated samples and a focus on the phase separation would be an interesting avenue for future experimental work since the result has implications on the long-term stability of the samples.

\bibliography{ref}